# An Overview of Researches on Laser Ion Acceleration Using Mixed Solid Target and Single Ion Target

**Jiafei Lyu**[1], **Rui Yang**[1], **Yu Yang**[1]

[1]Tsinghua Shenzhen International Graduate School

**ABSTRACT**: The essay gives an overview on researches in the field of laser ion acceleration, focusing on two types of targets. There are many types of targets while they can all be divided into targets that apply single ion or multiple ions. Mixed solid targets are proven efficient in accelerating heavy ions and generate high-quality ion beams with energy divergence lower than 5%. Traditional methods like TNSA are mainly used to accelerate protons or heavy ions and there are still many spaces for modification and improvement. Applications of laser-driven ion beams are wide in fields like detector technology, cancer therapy and so on, which is promising and necessary.

## 1 INTRODUCTION

Ultra-intense laser-plasma interaction has been deemed to be a bridge between fundamental physics and technological progress [1]. This coupling "has opened avenues using multi-MeV proton and ion beams" [2] for varieties of important applications such as fast ignitor of laser fusion [3] and accelerator technology [4]. Because of these significant applications, there is a tendency that laser ion acceleration becomes the focus of intensive and fundamental studies of physics.

In this paper, a brief introduction of some researchers' results and vital progress in the laser ion acceleration area will be presented. This essay will put its concentration on what is the difference between laser ion acceleration with mixed solid target and single ion target as well as targets with multiple layers. Moreover, readers can dig deeper into the area by reading the references which is given in APA style listed at the end of the essay or search on the website of PRL and other professional websites that concern about plasma and laser physics. The existing problems in the field are also pointed out in throughout the essay, especially the conclusion part.

Carbon target is applied in the BOA (break-out afterburner) [5] and usually a titanium target or PMMA target which is rich in proton is utilized in TNSA [6]. Both of these methods are proven effective while it is universally acknowledged that diamond and titanium is very hard. It indicates that it is very difficult for human beings to process these materials into thin targets. In contrast, foil is high in atomic number and is of excellent ductility, meaning that it is rich in proton and easy to be processed into any target of any thickness, which is the reason why many methods like CES would apply a foil target.

Many scientists in Osaka University like Ryosuke Kodama have interest in the process of laser ion acceleration inside irradiated solid targets by applying neutron spectroscopy, researchers from INFN (Instituto Nazionale di Fisica Nucleare [1]), however, seem to have more enthusiasm in realizing ion acceleration with a narrow energy spectrum by nanosecond laser-irradiation of solid target, which have something in common with researchers from University of Messina (Italy) and University of Catania (Italy) [7]. Similar researches are also conducted in universities and institutes all over the world like University of New South Wales in Australia [8], University of Strathclyde in UK [9], Pakistan Institute of Nuclear Science and Technology and Optics Laboratories in Pakistan [10]. As to the pure ion foil target, needless to say, there exist plenty of papers or experiments that concerns about foil under different situations. Generally, the studies of laser ion acceleration by utilizing pure ion foil target are centralized in countries like China, Germany, America, Japan, Australia etc. Institutes and universities such as Institute of Physics of Chinese Academy of Sciences, Los Alamos National Laboratory in USA, Max-Born-Institut in Germany, Photo medical Research Center in Japan, Peking University in China, University of Strathclyde in UK, CLPU (Centro de Laseres Pulsados[2]) in Spain etc.

Mixed solid target is comparatively a new concept in recent decades, serving to achieving high-quality ion beams. Researchers from Shanghai Institute of Optics and Fine Mechanics performed an experiment in 2009 on ion acceleration with mixed solid targets with circularly polarized lasers [11]. Also, previously, there are many theoretical and experimental work that serves to pave the way for laser interaction with mixed solid target. Institutes and universities like Shanghai Institute of Optics and Fine Mechanics, Institute for Plasma Research of India, Los Alamos National Laboratory of USA, University of Nevada, Max-Planck-Institut fur Quantenoptik of Germany and so on are embarking on similar projects [11-20].

As the research topic is laser ion acceleration, naturally, there are generally three difficult problems to be solved in the essay: (1) How can we acquire high energy ion beams by applying laser ion acceleration? (2) How to accelerate particles so as to achieve the best result like more energy deposit and less energy divergence? (3) Where can laser ion acceleration be possibly applied?

---

[1] Italian, it means National Institute of Nuclear Physics

[2] Spanish, it means Pulsed Laser Center

## 2 MIXED SOLID TARGET

As we all know, heavy ions cannot be accelerated efficiently just like protons as their lower charge-mass ratio. While according to [11-15], we do can increase the velocity of ions and accelerate them efficiently with mixed solid targets. Here, the word "mixed" means that there are two species of ion plasmas in the target during the interaction. Researches find that the velocity of heavy ions increases with the rise of the proton proportion. The theory as well as experiments are based on the acceleration of mixed cold targets with gaussian type pulse and ultrathin foil.

Considering one-dimensional particle in cell (1D PIC) situation where the ratio of charge-mass is fixed with the parameters set as follows according to [11,12]:

the wavelength of CP laser pulse: $\lambda = 1\mu m$

initial density: $n_e = 10$

the initial region that is occupied by targets is between $x = 64\lambda$ and $x = 72\lambda$

the laser amplitude rises from 0 to $a = 2$ in 5 times of laser period and then remains constant keep the ratio $\dfrac{n_{i1}}{n_{i2}} = 4$ constant reference [11] derives the process of the acceleration of heavy-ion and import the following figure at two instants $t = 80T$ and $t = 180T$:

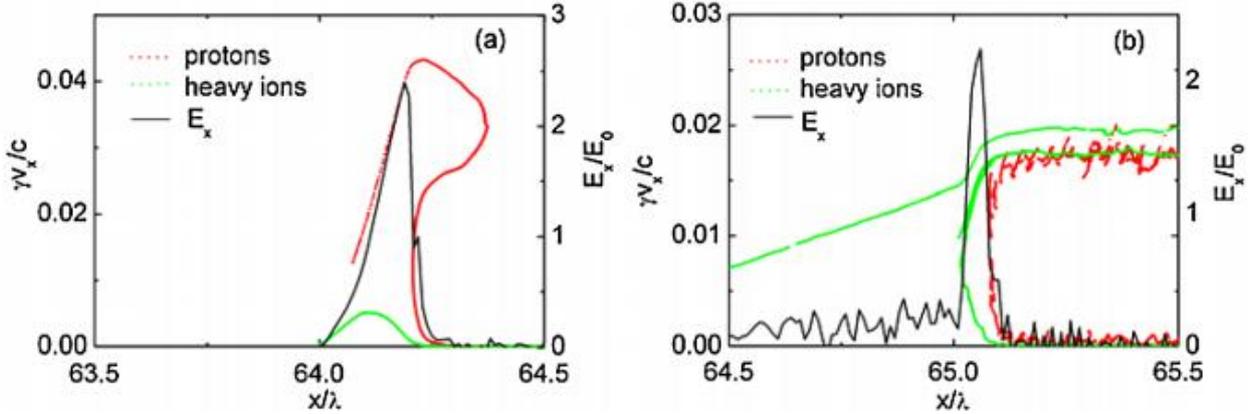

FIG. 1. Phase space of protons (red dots) and heavy ions (green dots) and the electric field distribution (black solid line) for the proton-dominated plasmas at (a) $t = 80T$ and (b) $t = 180T$.

It shows vivid in FIG. 1 that there exist two stages of different acceleration: when $t = 80T$, protons are accelerated efficiently up to a considerable speed by the electrostatic field while the heavy ions almost do not get much accelerated at all, which can even be seen as immobile due to their larger inertia. Light ions charge faster than the heavy one, meaning that the heavy ions do not have enough time to respond the sudden change of the field. But when it turns to $t = 180T$, things are different. It shows clearly in FIG. 1 (b) that heavy ions almost overtake protons and it seems that the reflected protons and heavy ions almost possess the same velocity. When acceleration time reaches to the magnitude of hundreds of pulse periods, the interaction can be seen as stable enough and heavy can be fully accelerated. It is emphasized that

*The heavier ion acceleration is more efficient in the mixed plasmas than in the pure-heavy-ion plasma.* [11,12,15].

Contrary to protons who have smaller mass, heavy ions are accelerated by a "much more partial electrostatic field", leading to the enormous changes in their velocities [11-15]. Also, we can find from FIG. 1 that heavy ions are driven by lighter ions indirectly which explains why light ions are slightly ahead of heavy ions. The final velocities of light and heavy ions are nearly the same as the accelerating electric field belong to the same self-consistent electrostatic shock system, which can also be seen from the equation (1), thus the velocities are determined only by the inherent parameters of the laser and plasma.

We are interested in the exact influence of charge-mass ratio and heavy ion proportion on the result of the acceleration theoretically and experimentally.

With the definition of two corresponding variables: $\alpha = n_{e1}/n_{e2}$, $\beta = (Z_1/A_1)/(Z_2/A_2)$, a vital equation can be derived [11,15]:

$$\frac{v_i}{c} \approx 2\sqrt{1+\frac{\alpha(\beta-1)}{\alpha+\beta}}\sqrt{\frac{Z_2}{A_2}\frac{m_e}{m_p}\frac{n_c}{n_e}}a \qquad (1)$$

where $n$ is density, $m$ is mass, $Z$ is proton number, $A$ is atomic number and $a$ is during period of the process.

The result of the simulation shows that such shock acceleration mechanism exists for a wide range of possible parameters. Moreover, to raise the energy of heavy ions effectively, it is a must that the proportion of heavy ions be much lower than that of light ions. But it is also important that the proportion of heavy ions should not be too small, which may contribute to the failure of acquiring energy increase at the second stage of acceleration.

The figure of different results of simulation and experiments under different values of $\alpha$ and $\beta$ are shown below where points stand for experiments and solid-line curve for simulation [11,15]:

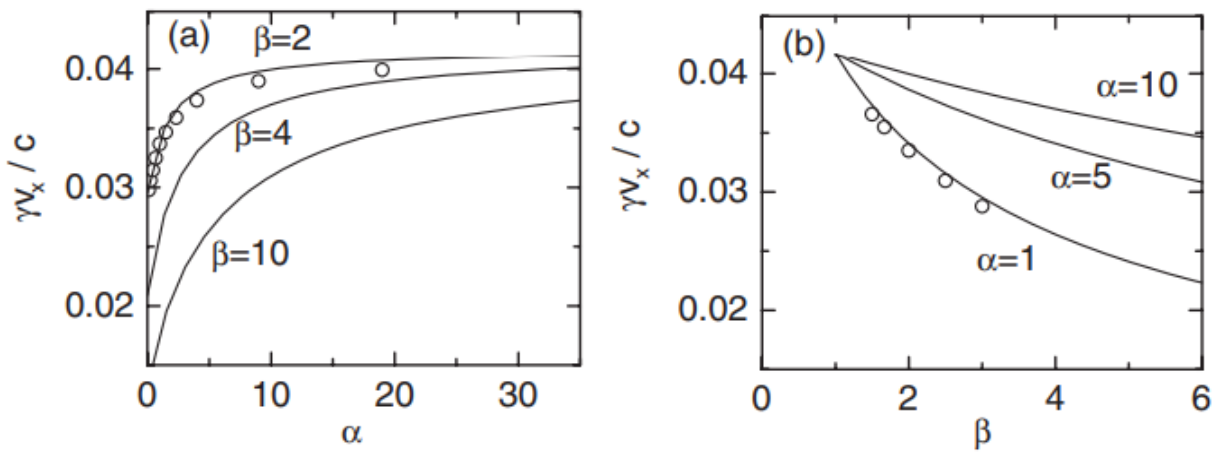

FIG. 2. Reflected heavy-ion momentum versus (a) $\alpha$ and (b) $\beta$ from simulations (circles) at $t = 150T$.

We can see easily from FIG 2 (a) that the proportion of heavy ions should be much less than that of the light ions so as to raise the energy of the heavy ions efficiently as the curve declines with the decrease of $\alpha$. FIG 2 (a) also illustrates us of the influence of the ratio of heavy ions on its acceleration which can be seen from the curve with parameter $\beta = 10$. From FIG. 2 (b), it is clear that the energy declines with the increasement of $\beta$ which indicates that we should not choose a low value of $\beta$. Moreover, we can find that the decline of $\beta$ is not sharp when is $\alpha$ large.

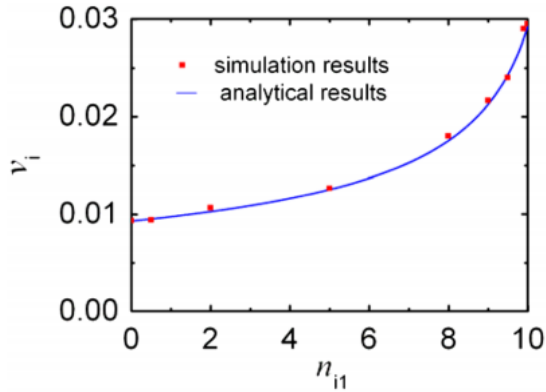

FIG. 3. Profile of the ion velocity as the function of proton density. The red square denotes the simulation results and the blue line denotes the analytical results [11,15].

FIG. 3, however, show us that the acceleration effect of the heavy ions is better with a larger density ratio of the light and heavy ions, which is acceptable and intuitive [11]. While results of similar but simpler researches [16,17] do not accord with the result shown above and it is maybe a flaw in the method which needs to be checked in future researches.

It has to be pointed out that the previous analysis is based on the hypothesis that the laser pulse is flattop which means that the accelerated ions are monoenergetic. However, in most practical cases the laser is roughly Gaussian and maybe plasma density is not always a constant in space, which probably leads to the ions not being monoenergetic. As a solution to the issue, a sandwich target with a thin mixed layer between two light ion layers just as FIG. 4 shows is employed [11,12,15].

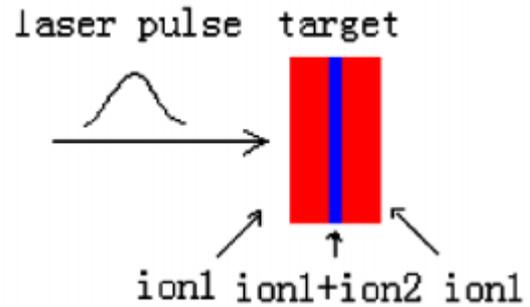

FIG. 4. Sandwich target scheme from the normal mixed target (red solid) and sandwich target (blue solid) at $t = 240T$ for a Gaussian laser pulse with peak amplitude $a = 4$ and FWHM of $22T$.

As is mentioned before, the pulse employed is gaussian type. Therefore, when the rising edge of the pulse reaches the target, it can only interact with ion1 in the front layer with little energy deposit, same as the right part of the pulse, and when the most intense part of the pulse interacts with the target, it almost locates in the middle part. The middle layer is thin to ensure that the interaction time is short enough to keep the laser intensity constant, resulting in "a more monoenergetic heavy-ion bunch" [11,14,15,20].

There is nothing ambiguous about the fact that mixed solid target is a promising direction for accelerating heavy ions. Currently, by applying the method of mixed solid target, a laser intensity of $5\times 10^{19}$ W/cm², heavy ion beams with energy divergence of 5% at the longitudinal kinetic energy of $E_x$ ~58 MeV with the total charge of ~$10^{-10}$ C can be obtained [11,12,15,18-20]. It shows that such method is practicable and applicable. In addition, high laser intensity and low temperature within a certain range are appropriate for accelerating heavy ions. To sum up, the method is so effective and practical that it has been proven that under the same laser condition, the method can raise the energy of heavy ions by nearly 100% [11-15,18-20]. The method is of great importance because by applying such method, the quality as well as the energy and space divergence of the heavy-ion beam can be improved tremendously. Moreover, the parameters of laser and plasma applied in the simulation and calculation can be easily achieved in the laboratory, which indicates an efficient way of producing high-quality heavy-ion beams.

It is also meaningful to introduce the general acceleration mechanism of mixed solid target. CES method is a typical example. CES is the abbreviation of collision-less electrostatic shock, originating in 1960s by a number of researchers like Moiseev, Montgomery and so on [38]. It has been proven to be an effective way of accelerating heavy ions theoretically and experimentally. The principle of CES is delicate: when an ultra-intense ultra-short pulse interact with mixed solid target, electrons would be expelled out of the target to form an electron spike just like methods mentioned above, which would decelerate incident ion beam. The delayed ions that do not have enough time to respond the pulse would then be accelerated by the electrostatic field established between ion spike and electron spike [38,39]. The unperturbed plasma ions that exist between the moving spikes would be reflected by the electric field and acquire nearly the same velocity regardless of the mass. Thus, an electrostatic shock propagating into a plasma is formed. During the reflection, the light ions and the heavy ions can get the same speed, resulting in the higher energy in heavy ions thanks to their heavier mass [15,38,39].

The method of mixed solid target mentioned above has some overlap with CES to some degrees. By applying CES, it has been reported that we can achieve helium ion beams of ~ 13 MeV with energy spread of about 7% with the intensity of laser up to ~ $10^{20}$W/cm² [39]. It is a relatively good result as we can get high energy and nearly high-quality heavy ion beams.

It is pleasing that the method does provide us with practical solutions to the difficult problems of how to generate high-quality ion beams and how to accelerate ions. But what distresses us is that the acceleration of protons is negatively influenced and there is almost no possibility for us to get high-quality proton beams which has bad energy and space divergence in mixed solid target method. Furthermore, currently it is relatively difficult to increase the laser density, causing the slowness of velocities of the accelerated ions [11,15,19,38,39]. The primary difficulty of the theory and experiments lie in how to choose parameters like heavy-ion proportion of mixed solid target. Also, to put forward with such fascinating idea of mixed solid target itself is of full difficulty.

There are many possible and promising applications of such method as we can acquire high-quality monoenergetic ion beams like proton beams or heavy ion beams. It can be applied in fast ignition of inertial confinement fusion [3], cancer and tumor therapy [21-23], progress in spectrometers and detectors or other relative equipment [24,25] and so on.

## 3 SINGLE ION TARGET

In this part, some traditional methods for accelerating ions utilizing single ion target as well as multiple layer targets will be overviewed. Traditional methods consist of TNSA, BOA and DCE, other methods like RPA would be neglected due to the pure fact that they have been covered in the course.

**a. TNSA**

TNSA (target normal sheath acceleration) is put forward by Wilks, Langdon and Cowan in 2001 and it has been proven to be effective in accelerating protons up to highest energy of nearly 60 MeV [6,26]. The acceleration mechanism of TNSA is presented in FIG. 5 [27]:

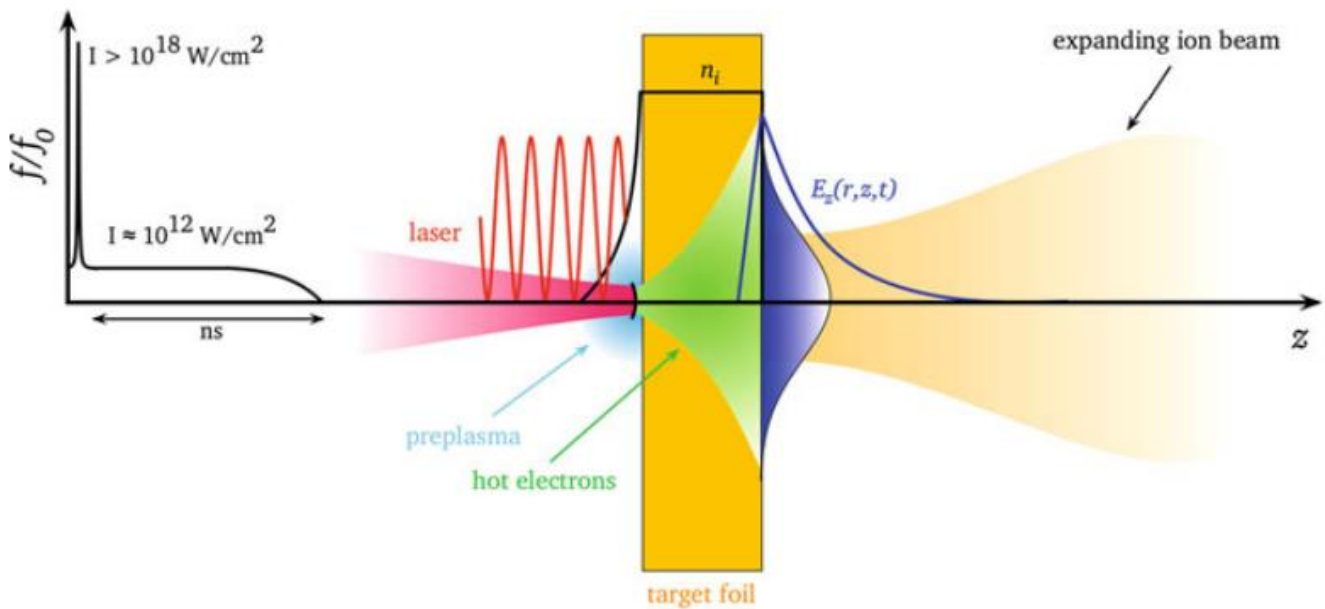

FIG. 5. The acceleration mechanism of TNSA

The target applied is single ion target. When the laser pre-pulse reaches the front side of the thin target foil, it creates a pre-plasma and then the main pulse interacts with the plasma, accelerating electrons forwardly with the ultimate kinetic energy up to magnitudes of MeV (generally 5 ~ 50 MeV) with ~ $10^{20}$ W/cm$^2$ lasers. Then the high energy electrons would collide with background materials, resulting in the increasement in the divergence of current. Afterwards, a highly dense sheath is formed along with the electric field due to the charge separation between electrons and ions, indicating that ions would be accelerated in the sheath electric field efficiently as the intensity of the field can reach magnitude of TV/m.

TNSA possesses the ability of accelerating all kinds of ions theoretically and the requirements of such method can be easily satisfied in the lab. While there are many flaws in the method, on one hand, the energy divergence of ion beams can reach 100% with a Maxwell distribution, which is contradictory to our expectation [28]. On the other hand, at the current stage, the energy conversion efficiency is very low. Though we can acquire approximate 70 MeV high energy protons but the number of high energy particles are few, which equals to say that such beams cannot be applied in reality, for example, tumor therapy.

Some scientists modify the TNSA method with a double layer target: one layer is a comparatively thick layer composed of plateau sub-ordinal material and the other ultrathin layer composed of materials with low atomic number [29,30]. Experiments and simulations prove that such modification can enhance the energy spread of ion beams down to ~17%, but still not efficient enough. It is still a difficult problem to solve in TNSA.

**b. BOA**

L.YIN came up with the idea of BOA in 2006 to generate carbon beams with GeV energy magnitude through the interaction of ultra-intense linear polarized laser and ultra-thin target [5,28,31]. The typical size of target is 100 nm and the typical energy of ion beams are 1~2 GeV which is very attractive and earthshaking. BOA also applies single ion target and the acceleration phase of BOA can be divided into 4 stages according to [28,31]. At first, the acceleration follows TNSA acceleration mechanism where hot electrons generated by the mechanism formed a sheath where the target is mainly made up with cold electrons. Then, more and more cold electrons are turned into hot electrons which enhances the intensity of the electric field. Electrons would move with relativistic velocity and drive the ions to be accelerated. Finally, the pulse passed through the target with ion beams of diverse energy.

It is obvious that the most beneficial part of BOA is ion beams with ultra-high energy as currently there is nearly no efficient ways to generate GeV ion beams. While there are also some drawbacks, BOA is demanding to targets (10~500 nm) and the energy spread of BOA is also big (10%~20%) compared to previous methods. Similar to TNSA, double-layer target is also applied to modify the method [32].

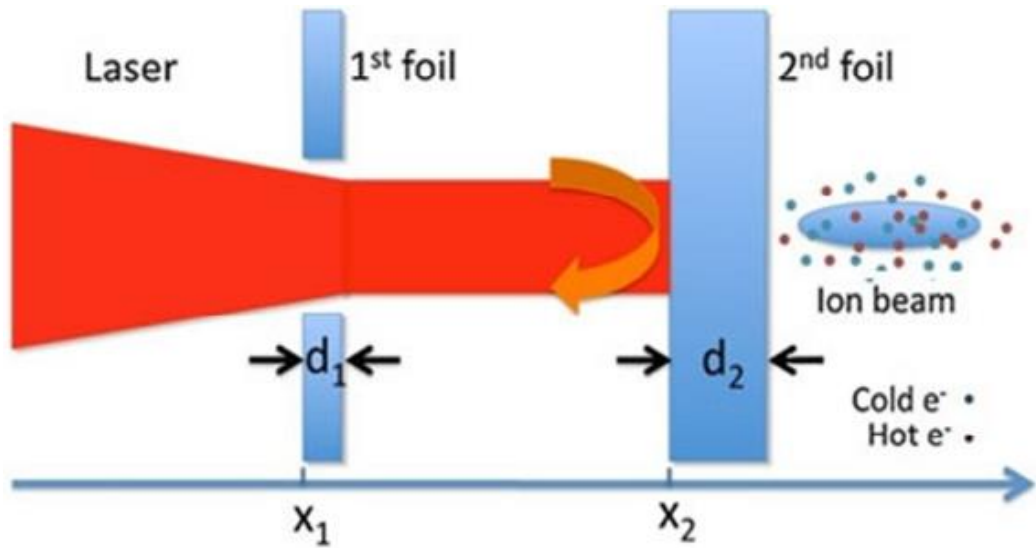

FIG. 6. The schematic diagram of double-layer BOA acceleration [28,32].

With the addition of a second foil with thickness and density that is properly chosen as is shown in FIG. 6, the energy divergence can be reduced to about 7% [32,33]. Another vital function of the second foil target is to prevent laser pulse from further heating electrons as well as provide a container for the exchange of cold electrons and hot electrons to ensure the balance and equilibrium of the system. BOA is an efficient and practicable way to accelerate ions overall.

**c. DCE**

DCE stands for directed coulomb explosion. It is a relatively old idea for acceleration with a single ion target, originating in 1967 by Fleischer, Price and Walker naming ion explosion [34]. The principle of DCE is very simple, the intense ultra-short laser ($>10^{16}$ W/cm²) interacts with the irradiated target, expelling electrons and establishing an ion layer. A strong electrostatic field is formed after the interaction which plays a critical role in accelerating protons. Then, protons would undergo a "coulomb explosion" because of the excess of positive charges, expanding like clouds from a bomb explosion in the laser propagation direction [35]. The coulomb explosion would affect the distribution and homogeneity of the electrostatic field and the electron layer. With the application of DCE, it is not difficult to acquire ion beams with 50 MeV and the energy spread 20% ~ 50%.

While there are also many problems with the method, such as the accelerated protons are not satisfying as we desire for high energy ion beams and the vast energy divergence is disappointing. In order to handle the problem, two-layer target is employed [35-37].

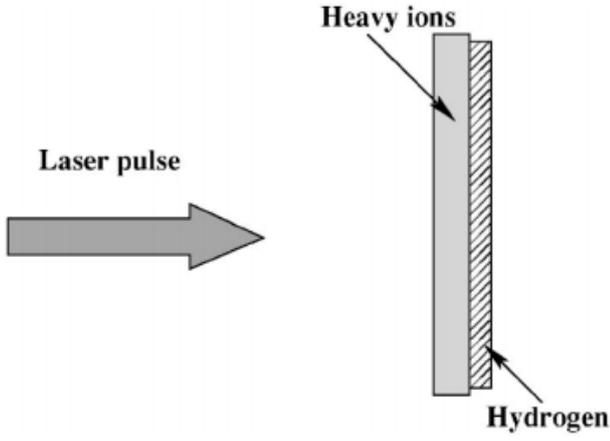

FIG. 7. Schematic diagram of two-layer target system [36].

The first layer is composed of intense heavy ions and the second layer, which is very thin, is made up of ionized low-density hydrogen. When the ultra-short ultra-intense laser pulse interact with the front layer, electrons escape from the layer with the charged layer of heavy ions left [37]. The electrostatic field established in the target is strong enough to accelerate and improve the unipotency of protons efficiently. The dynamics of coulomb explosion, however, are ignored as the ion mass are much larger than that of proton.

With the modification, we can get proton beams with more than 100 MeV, approximate $10^7$ protons generated and energy divergence less than 5% by utilizing a laser of 500 TW with the simulation result of 111 MeV and energy spread of 2.8% [37]. Also, we can get ion beams with energy ranging from 140 to 210 MeV and energy spread at about 3% for a Gaussian beam and a flat-top beam of $10^{21}$ W/cm² [35]. In summary, we can produce comparatively high-quality proton beams by applying the modified DEC method with energy ~ 100 MeV and energy spread ~ 5%. But the quality of the ion beams still needs to be improved.

## 4 CONCLUSIONS

In recent decades, there are more and more needs for monoenergetic ion beams with high energy like magnitude of MeV or more in fields like medical therapy, accelerator advance and fast ignition of laser-driven fusion [3,15,18,22,23,40]. If we can get absolute monoenergetic ion beams, then the question (2) that is aroused in introduction part is completely solved. While unfortunately, we can not erase the energy spread of ion beams at the current stage and it is also impossible to erase the energy divergence due to the randomness of interaction between pulse and plasma. Furthermore, if we apply a detector to detect the accelerated ion beams despite of it is made up of positive ions or negative ions, we get an energy spectrum with energy spread due to the statistical randomness of interaction between beams and detector medium and all detectors have their inherent energy resolution, i.e. FWHM [41]. We can also define the energy resolution of ion beams out of the target which is actually called as energy spread. The definition of energy spread is as follows:

$$\eta = \frac{\Delta E}{E} \qquad (2)$$

in equation (2) $\Delta E$ is the spectral width at half the peak height. We expect less energy spread to improve the quality of the beams. In TNSA, the phenomenon is devastating where the energy spread is 100% while in BOA, we can acquire energy spread of about 17% and 20% ~ 50% in DCE. In contrast to those methods, in mixed solid target and CES, we can acquire energy spread of less than 5% which is still not enough for practical application, for example, in cancer therapy, energy divergence of less than 1% is required [21-23]. The modification of these methods is usually conducted through applying two-layer target with different properties in two layers. After modification, the energy spread can be sharply reduced to a small number while there still remains lots of work to be done in the field of laser ion acceleration.

As to the first question that how to acquire high energy ion beams, different methods give different maximum energy of ion beams. In TNSA, we can acquire ~50 MeV high energy ion beams and ~ 1 GeV in BOA, ~ 100 MeV in DCE, ~ 58 MeV in mixed solid target. Such comparison is not valid as some of the methods are applied to accelerate heavy ions like mixed solid target as heavy ions are far more difficult to be accelerated than light ions like protons while others are employed to generate high energy proton beams. Actually, compound targets were put in use in the course of laser-driven ion acceleration, but they were often used as a background to increase "the gradient of the space-charge field" for light ions like protons [15,42].

The third question concerning about the application fields of laser ion acceleration has been answered during the introduction to different targets and acceleration mechanisms. It is noteworthy that PIC (particle in cell) simulations are often applied to laser ion acceleration as a test of theory, which is powerful and appropriate when developing new methods for acceleration [3-15,18-20,26-33,41].

In summary, the essay introduces two different targets: target that contain only single type of ion and target that contain two types of ions. single ion targets are effective in accelerating light ions like protons and electrons while mixed solid target is applied to accelerate heavy ions. Mixed solid target is still not mature and remain lots of problems to solve like how to improve the maximum energy of ion beams and how to accelerate more heavy ion particles compared to the well-developed single ion target. Different targets are chosen under different situations according to the requirements of the researchers and sometimes multiple layers of targets are used. The commonly used material for targets is foil at the current stage [12,13,17]. The targets tend to be made more and more thin to magnitude of nanometer. It is easier for electrons in ultra-thin targets to be expelled and can improve the unipotency of ion beams [8,40]. Meanwhile, to ensure the sufficient interaction between targets and laser pulse, it is necessary that the thickness of targets to be close to the wavelength of laser pulse.

The thickness of targets can reach the magnitude of $10^{-9}$ m at present [31]. Compared to foil, it is difficult for titanium to be processed into thin targets while it is reported that Schwoerer et al. who come from Germany have succeeded in producing titanium targets of 5 μm [43]. Thinner a target is, harder will it be to process it. Anyway, the technology of processing targets as well as the methods of accelerating ions is getting more and more mature.

Nevertheless, there are still many problems in the laser-driven ion acceleration. The mechanism and location of ion acceleration as well as where do the observed ions come from are still under debate despite the fact that some experiments have shown that the acceleration process of protons happens at the back surface of the target [36,44]. Also, different targets composed of different materials with different thickness have different influence on the acceleration process which should be estimated in future researches. Meanwhile, it is necessary to consider the stability of the target and the possible deformation of the thin and fragile target when interacting with ultra-intense ultra-short laser pulse. Researches have revealed that the lateral instability, such as Weibel instability and Rayleigh-Taylor-like instability, and the deformation of target become more and more important with the simulation of multi-dimensional particles [28,45,46]. Under such condition, it is vital to suppress the deformation of the target and the lateral instability in order to acquire high-quality ion beams. The issues mentioned before are still under consideration and tested which appeals for practical solutions.

In general, laser ion acceleration method is promising and practicable both experimentally and theoretically. But there is still a long way to go to fully meet the requirement of practical application through the utilization of mixed solid target and single ion target as well as different acceleration mechanisms.

## 5 REFERENCES

[1] Dyson, A. . (1999). Ultra-intense laser-plasma interactions.

[2] Youssef, A., Kodama, R., & Tampo, M. . (2006). Investigation of laser ion acceleration inside irradiated solid targets by neutron spectroscopy. *Physics of Plasmas, 13*(3), 030701.

[3] Snavely, R. A. , Key, M. H. , Hatchett, S. P. , Cowan, T. E. , Roth, M. , & Phillips, T. W. , et al. (2000). Intense high-energy proton beams from petawatt-laser irradiation of solids. *Physical Review Letters, 85*(14), 2945-2948.

[4] Esarey, E. , Sprangle, P. , Krall, J. , & Ting, A. . (1996). Overview of plasma-based accelerator concepts. *Plasma Science IEEE Transactions on, 24*(2), 252-288.

[5] Jung, D., Yin, L., Gautier, D. C., Wu, H. C., Letzring, S., & Dromey, B., et al. (2013). Laser-driven 1 GeV carbon ions from preheated diamond targets in the break-out afterburner regime. *Physics of Plasmas, 20*(8), 1775.

[6] Ceccotti, T., Lévy, A., Réau, F., Popescu, H., Monot, P., & Lefebvre, E., et al. (2008). TNSA in the ultra-high contrast regime. *Plasma Physics & Controlled Fusion, 50*(12), 2165-2168.

[7] Torrisi, L., Cutroneo, M., Calcagno, L., Rosinski, M., & Ullschmied, J. (2014). TNSA ion acceleration

at $10^{16}$ W/cm$^2$ sub-nanosecond laser intensity.

[8] Hora, H. (2012). Fundamental difference between picosecond and nanosecond laser interaction with plasmas: Ultrahigh plasma block acceleration links with electron collective ion acceleration of ultra-thin foils. *Laser and Particle Beams, 30*(2), 325-328.

[9] Mckenna, P. , Ledingham, K. W. D. , & Robson, L. . (2006). *Laser-Driven Ion Acceleration and Nuclear Activation. Lasers and Nuclei*. Springer Berlin Heidelberg.

[10] Abbasi, S., Hussain, M., Ilyas, B., Rafique, M., Dogar, A., & Qayyum, A. (2015). Characterization of highly charged titanium ions produced by nanosecond pulsed laser. *Laser and Particle Beams, 33*(1), 81-86.

[11] Zhang, X., Shen, B., Ji, L., Wang, F., Jin, Z., & Li, X., et al. (2009). Ion acceleration with mixed solid targets interacting with circularly polarized lasers. *Phys. Rev. Spec. Top.-Accel. Beams, 12*(2), 64-64.

[12] Ji, Liangliang, Shen, Baifei, Zhang, Xiaomei, Wang, Fengchao, Jin, Zhangyin and Li, Xuemei. (2009). Single energy heavy ion beam generated by laser guided electrostatic shock wave. *Progress in laser and Optoelectronics*, *46*(2).

[13] Yu, T. P. , Pukhov, A. , Shvets, G. , & Chen, M. . (2010). Stable laser-driven proton beam acceleration from a two-ion-species ultrathin foil. *Physical Review Letters, 105*(6), 065002.

[14] Dhareshwar, L. , & Chaurasia, S. . (2008). Laser plasma interaction in solid metal, mixed metal alloy and metal Nano-particle coated targets. *Journal of Physics: Conference Series, 112*(3), 032050.

[15] Ji, L. , Shen, B. , Zhang, X. , Wang, F. , Jin, Z. , & Li, X. , et al. (2008). Generating monoenergetic heavy-ion bunches with laser-induced electrostatic shocks. *Physical Review Letters, 101*(16), 164802.

[16] Albright, B. J. , Yin, L. , Hegelich, B. M. , Bowers, K. J. , Kwan, T. J. T. , & Fernández, J. C. (2006). Theory of laser acceleration of light-ion beams from interaction of ultrahigh-intensity lasers with layered targets. *Physical Review Letters, 97*(11), 115002.

[17] Hegelich, M. , Karsch, S. , Pretzler, G. , Habs, D. , Witte, K. , & Guenther, W. , et al. (2002). MeV ion jets from short-pulse-laser interaction with thin foils. *Physical Review Letters, 89*(8), 085002.

[18] Wang, F. , Shen, B. , Zhang, X. , Jin, Z. , Wen, M. , & Ji, L. , et al. (2009). High-energy monoenergetic proton bunch from laser interaction with a complex target. *Physics of Plasmas, 16*(9), 093112.

[19] Korzhimanov, A. V. , Efimenko, E. S. , Golubev, S. V. , & Kim, A. V. . (2012). Generating high-energy highly charged ion beams from petawatt-class laser interactions with compound targets. *Physical Review Letters, 109*(24), 245008.

[20] Sinha, & Ujjwal. (2012). Self-consistent model for ponderomotive ion acceleration of laser irradiated two species dense target plasmas. *Physics of Plasmas, 19*(4), 43104-0.

[21] Schardt, D. , Thilo Elsässer, & Schulz-Ertner, D. . (2010). Heavy-ion tumor therapy: physical and

[22] Jakel, O. , Schulz-Ertner, D. , Karger, C. P. , Nikoghosyan, A. , & Debus, J. . (2003). Heavy ion therapy: status and perspectives. *Technology in Cancer Research & Treatment, 2*(5), 377-387.

[23] Torikoshi, M. . (2006). Heavy-ion cancer therapy. *Laser Physics, 16*(4), 654-659.

[24] Turchetti, G. , Malka, V. , & Giovannozzi, M. . (2010). Ions acceleration with high power lasers: physics and applications. *Nuclear Inst & Methods in Physics Research A, 620*(1), vii-viii.

[25] Cutroneo, M. , Musumeci, P. , Zimbone, M. , Torrisi, L. , La Via, F. , & Margarone, D. , et al. (2013). High performance sic detectors for mev ion beams generated by intense pulsed laser plasmas. *Journal of Materials Research, 28*(01), 87-93.

[26] Wilks, S. C. , Langdon, A. B. , Cowan, T. E. , Roth, M. , Singh, M. , & Hatchett, S. , et al. (2001). Energetic proton generation in ultra-intense laser–solid interactions. *Physics of Plasmas, 8*.

[27] Roth, M. , & Schollmeier, M. . (2013). *Ion Acceleration: TNSA. Laser-Plasma Interactions and Applications*. Springer International Publishing, 311.

[28] Weiquan Wang. (0). *A study of Ion acceleration of thin film target driven by super intense laser*. (Doctoral dissertation).

[29] Esirkepov, T. Z. , Bulanov, S. V. , Nishihara, K. , Tajima, T. , Pegoraro, F. , & Khoroshkov, V. S. , et al. (2002). Proposed double-layer target for the generation of high-quality laser-accelerated ion beams. *Physical Review Letters, 89*(17), 175003.

[30] Gu, Y. J. , Kong, Q. , Kawata, S. , Izumiyama, T. , Li, X. F. , & Yu, Q. , et al. (2013). Enhancement of proton acceleration field in laser double-layer target interaction. *Physics of Plasmas, 20*(7), 22.

[31] Yin, L. , Albright, B. J. , Hegelich, B. M. , & Fernández, J. C. (2006). GeV laser ion acceleration from ultrathin targets: the laser break-out afterburner. *Laser & Particle Beams, 24*(2), 291-298.

[32] Huang, C. K. , Albright, B. J. , Yin, L. , Wu, H. C. , Bowers, K. J. , & Hegelich, B. M. , et al. (2011). Improving beam spectral and spatial quality by double-foil target in laser ion acceleration. *Physical Review Special Topics - Accelerators and Beams, 14*(3), 031301.

[33] Huang, C. K. , Albright, B. J. , Yin, L. , Wu, H. C. , Bowers, K. J. , & Hegelich, B. M. , et al. (2011). A double-foil target for improving beam quality in laser ion acceleration with thin foils. *Physics of Plasmas, 18*(5), 056707.

[34] Fleischer, R. L. , Price, P. B. , & Walker, R. M. . (1965). Ion explosion spike mechanism for formation of charged-particle tracks in solids. *Journal of Applied Physics, 36*(11), 3645-3652.

[35] Bulanov, S. , Brantov, A. , Bychenkov, V. , Chvykov, V. , Kalinchenko, G. , & Matsuoka, T. , et al. (2008). Accelerating monoenergetic protons from ultrathin foils by flat-top laser pulses in the directed-coulomb-explosion regime. *Physical Review E (Statistical, Nonlinear, and Soft Matter Physics), 78*(2), 26412-0.

[36] Fourkal, E. , Velchev, I. , & Ma, C. M. . (2005). Coulomb explosion effect and the maximum energy
**14** / **15**

of protons accelerated by high-power lasers. *Physical Review E, 71*(3), 036412.

[37] Morita, T., Bulanov, S. V., Esirkepov, T. Z., Koga, J., & Kando, M. (2011). Directed coulomb explosion effect on proton acceleration by an intense laser pulse from a double-layer target. *Physics*.

[38] Forslund, D. W. , & Shonk, C. R. . (1970). Formation and structure of electrostatic collision less shocks. *Physical Review Letters, 25*(25), 1699-1702.

[39] Wei, M. S. , Mangles, S. P. D. , Najmudin, Z. , Walton, B. , Gopal, A. , & Tatarakis, M. , et al. (2004). Ion acceleration by collision less shocks in high-intensity-laser–under-dense-plasma interaction. *Physical Review Letters, 93*(15), 155003.

[40] Yan, X. Q. , Tajima, T. , Hegelich, M. , Yin, L. , & Habs, D. . (2010). Theory of laser ion acceleration from a foil target of nanometer thickness. *Applied Physics B: Lasers and Optics, 98*(4), 711-721.

[41] Knoll, & Glenn, F. . (1989). *Radiation detection and measurement / Glenn F. Knoll. Radiation detection and measurement*. Wiley.

[42] Hegelich, B. M. , Albright, B. J. , Cobble, J. , Flippo, K. , Letzring, S. , & Paffett, M. , et al. (2006). Laser acceleration of quasi-monoenergetic MeV ion beams. *Nature, 439*(7075), 441.

[43] Schwoerer, H. , Pfotenhauer, S. , Jäckel, O, Amthor, K. U. , Liesfeld, B. , & Ziegler, W. , et al. (2006). Laser plasma acceleration of quasi-monoenergetic protons from microstructured targets. *Nature, 439*(7075), 445-8.

[44] Snavely, R. A., Key, M. H., Hatchett, S. P., Cowan, T. E., Roth, M., & Phillips, T. W., et al. (2000). Intense high-energy proton beams from petawatt-laser irradiation of solids. *Physical Review Letters, 85*(14), 2945-8.

[45] Pegoraro, F. , & Bulanov, S. V. . (2009). Stability of a plasma foil in the radiation pressure dominated regime. *The European Physical Journal D, 55*(2), 399-405.